%% file: main.tex
\documentclass[sigconf]{acmart}
\AtBeginDocument{%
  }

\copyrightyear{2025}
\acmYear{2025}
\setcopyright{rightsretained}
\acmConference[RecSysChallenge '25]{RecSys Challenge 2025}{September 22, 2025}{Prague, Czech Republic}
\acmBooktitle{RecSys Challenge 2025 (RecSysChallenge '25), September 22, 2025, Prague, Czech Republic}
\acmPrice{}
\acmDOI{10.1145/3758126.3758132}
\acmISBN{979-8-4007-2099-4/25/09}

\usepackage{makecell}
\usepackage{multirow}
\usepackage{graphicx}
\usepackage{enumitem}
\usepackage{hyperref}

\begin{document}

\title{Encode Me If You Can: Learning Universal User Representations via Event Sequence Autoencoding}

\author{Anton Klenitskiy}
\email{antklen@gmail.com}
\orcid{0009-0005-8961-6921}
\affiliation{
  \institution{Sber AI Lab}
  \city{Moscow}
  \country{Russian Federation}  
}

\author{Artem Fatkulin}
\email{artem42fatkulin@gmail.com}
\orcid{0009-0006-7105-6003}
\affiliation{
  \institution{Sber AI Lab}
  \city{Moscow}
  \country{Russian Federation}
}
\affiliation{
  \institution{HSE University}
  \city{Moscow}
  \country{Russian Federation}  
}

\author{Daria Denisova}
\email{duny.explorer@gmail.com}
\orcid{0009-0006-2918-7798}
\affiliation{
  \institution{Sber AI Lab}
  \city{Moscow}
  \country{Russian Federation}
}

\author{Anton Pembek}
\email{apembek@bk.ru}
\orcid{0009-0005-1757-3379}
\affiliation{%
  \institution{Sber AI Lab}
  \city{Moscow}
  \country{Russian Federation}
}
\affiliation{%
  \institution{Lomonosov Moscow State University (MSU)}
  \city{Moscow}
  \country{Russian Federation}
}

\author{Alexey Vasilev}
\email{alexxl.vasilev@yandex.ru}
\orcid{0009-0007-1415-2004}
\affiliation{
  \institution{Sber AI Lab}
  \city{Moscow}
  \country{Russian Federation}
}
\affiliation{
  \institution{HSE University}
  \city{Moscow}
  \country{Russian Federation}
}

\renewcommand{\shortauthors}{Klenitskiy et al.}

\begin{abstract}
    \input{content/0_abstract}

\end{abstract}



\maketitle

\section{Introduction}

\input{content/1_intro}

\section{Problem Statement}
\input{content/2_problem}

\section{Approach}
\input{content/3_approach}

\section{Results}

\input{content/4_results}

\section{Conclusion}

\input{content/5_conclusion}

\bibliographystyle{ACM-Reference-Format}
\bibliography{content/bibliography}

\end{document}

%% file: content/0_abstract.tex
Building universal user representations that capture the essential aspects of user behavior is a crucial task for modern machine learning systems. In real-world applications, a user’s historical interactions often serve as the foundation for solving a wide range of predictive tasks, such as churn prediction, recommendations, or lifetime value estimation. Using a task-independent user representation that is effective across all such tasks can reduce the need for task-specific feature engineering and model retraining, leading to more scalable and efficient machine learning pipelines. The goal of the RecSys Challenge 2025 by Synerise was to develop such Universal Behavioral Profiles from logs of past user behavior, which included various types of events such as product purchases, page views, and search queries.

We propose a method that transforms the entire user interaction history into a single chronological sequence and trains a GRU-based autoencoder to reconstruct this sequence from a fixed-size vector. If the model can accurately reconstruct the sequence, the latent vector is expected to capture the key behavioral patterns. In addition to this core model, we explored several alternative methods for generating user embeddings and combined them by concatenating their output vectors into a unified representation. This ensemble strategy further improved generalization across diverse downstream tasks and helped our team, ai\_lab\_recsys, achieve second place in the RecSys Challenge 2025.

%% file: content/1_intro.tex
\begin{figure*}[!ht]
  \centering
    \includegraphics[width=0.9\textwidth]{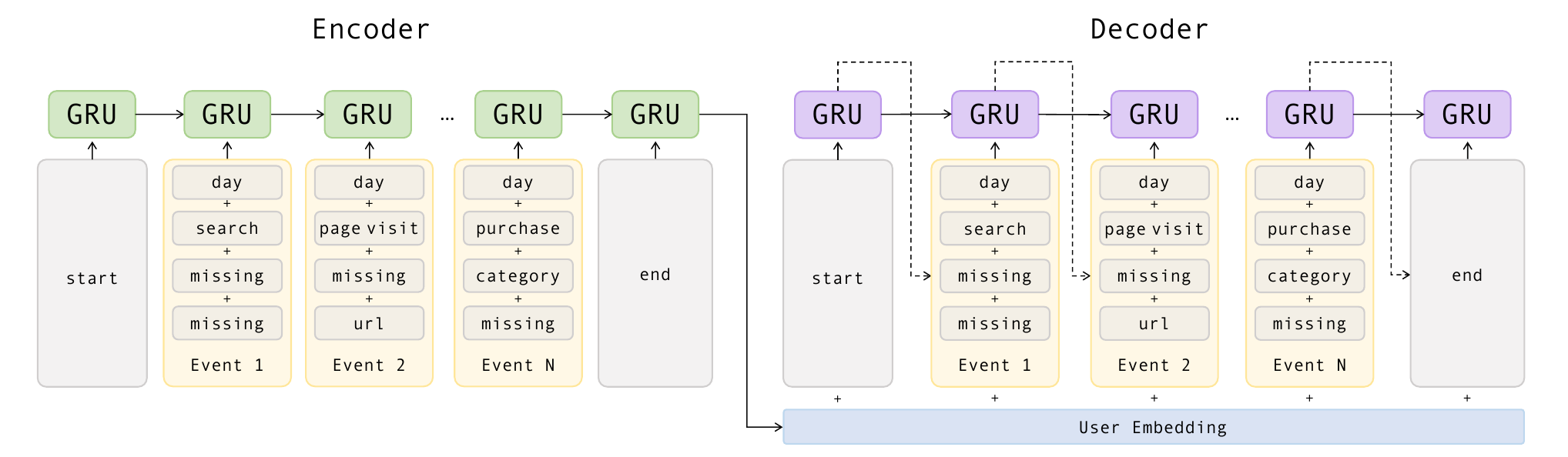}
  \caption{Example of GRU Autoencoder with four categorical fields (day index, event type, category, and url). Event embedding is a sum of individual embeddings. In the decoder, user embedding is added at each step.}
  \label{fig:gru_autoencoder}
  \Description{GRU Autoencoder.}
\end{figure*}

In many modern applications of machine learning, user behavior data plays a central role. Logs of user interactions are often used as the foundation for a variety of predictive tasks, including different recommendation scenarios, propensity prediction, churn prediction, user lifetime value prediction, and many others. Traditionally, each of these tasks is addressed with task-specific models and manually engineered features tailored to the target variable. However, such approaches are often difficult to scale and maintain across multiple use cases.

The RecSys Challenge 2025 by Synerise\footnote{\url{http://www.recsyschallenge.com/2025}} was designed to promote a unified approach to user behavior modeling. Participants were asked to create Universal Behavioral Profiles -- fixed-length user representations that encode essential aspects of each individual’s past interactions and are universally applicable across multiple predictive tasks. The data provided for building these profiles included various types of events: product purchases, adding or removing items from cart, page visits, and search queries. These representations were evaluated based on their performance across a diverse set of downstream tasks using a shared evaluation pipeline.

We propose an approach based on a sequence autoencoder that encodes a user's full interaction history into a single fixed-size vector. The model is trained to reconstruct the chronological sequence of user events from this fixed-size vector, forcing the latent representation to retain as much relevant behavioral information as possible. Each event in the sequence is represented as a set of categorical fields, which are passed through corresponding embedding layers. The sum of these embeddings is then fed into GRU (Gated Recurrent Unit~\cite{cho2014learning}) layers. For reconstruction, the model predicts each of these fields separately.

To further improve generalization, we construct an ensemble by incorporating several alternative embedding strategies and concatenating their outputs. These strategies include converting all data into text and using a language model as a feature extractor, training a transformer-based model to predict the next event, applying collaborative filtering methods (iALS and factorization machines), and manually engineering a set of features. We found that proper normalization of each embedding before concatenation is important for maximizing the ensemble performance.

Our team, ai\_lab\_recsys, took second place in the Challenge and first place on the academic leaderboard with a sum of scores across all downstream tasks of 4.7504 and a Borda count of 653. The implementation of our solution is available at GitHub\footnote{\url{https://github.com/antklen/recsys\_challenge\_2025}}.

%% file: content/2_problem.tex
\paragraph{Dataset.}

The Challenge aimed to develop universal behavioral profiles for users based on interaction data that encompass five types of events. All entries carry a client ID and a timestamp. The "product buy", "add to cart", and "remove from cart" events additionally record item metadata, comprising SKU (item ID), category, price bucket, and a textual embedding of the product name. The "page visit" events are associated with an encoded URL and therefore do not explicitly reveal page content. The final event type, "search query", includes embedding of the query text. All textual embeddings are derived from a large language model (LLM) and provided in quantized form. Table \ref{tab:statistic} summarizes the key dataset statistics.

\input{content/tables/dataset_statistic}

\paragraph{Tasks.}

User representations serve as input features for MLPs trained to predict user behavior over the next 14 days. The open tasks include churn prediction, a binary classification to determine whether a user will stop engaging; category propensity, a multi-label classification to forecast which of the 100 most common product categories a user will buy from; and product propensity, a multi-label classification to predict which of the 100 most frequently purchased SKUs a user is likely to purchase.

In addition, hidden tasks -- unveiled only after the end of the competition -- were designed to assess the models’ ability to generalize to unseen data and novel scenarios. They include conversion (hidden1), forecasting whether a user will make a purchase during the evaluation period; new product propensity (hidden2), predicting whether a user will purchase popular items that appear in the evaluation period but were not present during training; price propensity (hidden3), predicting price buckets a user is likely to engage.

\paragraph{Evaluation.}

All results are evaluated on a subset of 1 000 000 relevant clients. Notably, these clients tend to have much longer histories of page visits and search queries compared to all clients.

The primary performance metric is AUROC, while the final score for category and product propensity tasks also incorporates novelty and diversity: $\text{Score} = 0.8 \cdot \text{AUROC} + 0.1 \cdot \text{Novelty} + 0.1 \cdot \text{Diversity}$.
Models are ranked on separate leaderboards for each task based on respective scores. The organizers propose using the Borda count method to assess overall performance and generalization: a model ranked $k$-th out of $N$ in a task receives $N - k$ points. The final rankings are based on the total score across all tasks, favoring models with consistently strong performance.

%% file: content/tables/dataset_statistic.tex
\begin{table}[htbp]
  \caption{\textbf{Dataset statistics. Entities refers to URLs for "page visit" events and SKUs for "add to cart", "remove from cart", and "product buy" events.}}
  \label{tab:statistic}
  \centering
  \setlength{\tabcolsep}{4pt}
  \resizebox{\linewidth}{!}{
      \begin{tabular}{lrrrc}
        \toprule
        \textbf{Event type} &
        \textbf{\# Interact.} &
        \textbf{\# Clients} &
        \textbf{\# Entities} &
        \textbf{Avg.\ length} \\
        \midrule
        Page visit          & 156\,032\,014 & 18\,435\,078 & 12\,650\,786 & 8.46 \\
        Search query        & 10\,218\,831  & 1\,288\,493  & --           & 7.93 \\
        Add to cart         & 5\,674\,064   & 1\,887\,989  & 1\,178\,412  & 3.00 \\
        Remove from cart    & 1\,937\,170   & 552\,932     & 650\,078     & 3.50 \\
        Product buy         & 1\,775\,394   & 744\,980     & 519\,606     & 2.38 \\
        \bottomrule
    \end{tabular}
  }
\end{table}

%% file: content/3_approach.tex
\subsection{Sequence Autoencoder}

The intuition behind this approach is that training a model to compress a user’s entire interaction history into a single embedding and then reconstruct the original sequence from this representation is the most direct way to preserve as much information as possible. Since the resulting embedding is intended to be used across a variety of downstream tasks without prior knowledge of which aspects of the sequence are most relevant, maximizing the reconstruction fidelity serves as a strong general-purpose learning objective.

The first design choice concerns how to represent the input data. We begin by merging all available user interaction events from different sources into a single sequence, sorted by time. Each event is assigned an event\_type label indicating whether it is a "product buy", "add to cart", "remove from cart", "page visit", or "search query". In addition to the event type, each record includes several categorical fields: category, price (it can be treated as categorical since it was discretized into 100 buckets), SKU for product-related actions, and URL for page visits. To handle missing fields (e.g., no SKU for a page view or a search query), we assign a special "missing" value. For high-cardinality fields (SKU and URL), we retain only the 5,000 most frequent values and map all others to a shared "rare" value. We also encode temporal information by mapping each timestamp to a discrete day or week index within the observation window. This helps the model retain coarse-grained temporal structure, which is relevant for tasks like churn prediction, while discarding low-level temporal noise. To help the model recognize sequence boundaries, we append special "start-of-sequence" and "end-of-sequence" values to each sequence for every field.

Our model is illustrated in Figure~\ref{fig:gru_autoencoder}. It follows an encoder-decoder architecture~\cite{sutskever2014sequence} based on GRU recurrent neural network~\cite{cho2014learning}. Each categorical field is mapped to its own embedding space, and the final embedding of each event is computed as the sum of its field embeddings. The encoder consists of several stacked GRU layers, and the final user embedding is taken from the hidden state of the last GRU layer at the final sequence step. The decoder is also composed of several GRU layers. During decoding, user embedding is added to the input event embedding at each step. The model is trained with a sum of cross-entropy losses for each event field. We apply teacher forcing~\cite{williams1989learning} during training, feeding the decoder the original sequence rather than its own predictions.

We evaluate several variants of this architecture, comparing different subsets of fields used for reconstruction. \textit{GRU-AE week\_all} uses week index for temporal representation and all other categorical fields (event\_type, category, SKU, price, URL). \textit{GRU-AE all} is the same but without temporal information. \textit{GRU-AE day\_event\_type} is a simple version that uses only the day index and event\_type. These models are trained on the subset of relevant clients used for evaluation. We found that training on the full dataset did not improve performance. In addition to the models trained on all events, we also train a separate version of the autoencoder using only quantized textual embeddings of product names to incorporate this information into the final ensemble (\textit{GRU-AE sku\_text}). Each product is represented by a sequence of 16 discrete values, which can be treated as tokens. For each user, we take the last 4 interacted products and reconstruct a sequence of 64 tokens.

For \textit{GRU-AE day\_event\_type} we use a hidden size of 128, two GRU layers, and a dropout rate of 0.5. All other models are trained with a hidden size of 512, three GRU layers, and a dropout rate of 0.1. The maximum event sequence length is set to 128.

\input{content/tables/leaderboard}
\subsection{Other Approaches}

\subsubsection{iALS}

We apply iALS \cite{hu2008collaborative} to extract user propensity for specific products and categories. We train the model based on the user/category interaction matrix, selecting only "product buy" and "add to cart" events with weights of 3 and 1, respectively (\textit{iALS categories}). Similarly, we obtain embeddings based on the users' visited pages (\textit{iALS page visits}). The resulting representations from both models have a dimension of 64.

\subsubsection{LightFM}

We train a LightFM model \cite{kula2015metadata} using a user/SKU interaction matrix, incorporating "product buy", "add to cart", and "remove from cart" events with respective weights of 1, 0.7, and -0.5. As input features, we include product name embeddings and the average embedding of users' historical search queries. The resulting user representations have a dimensionality of 64.

\subsubsection{Transformer-based next event prediction}

We use two variants of a transformer-based model following the GPT-2 architecture~\cite{radford2019language}, trained from scratch to predict the next event in a sequence. The user embedding is extracted as the hidden state from the last transformer layer at the final step of the sequence. Unlike the autoencoder, which is trained to reconstruct the entire sequence, this model is optimized for next-step prediction, potentially focusing more on recent interactions and capturing shorter-term user intent.

The first variant (\textit{Next URL}) is trained on page visits data from relevant clients to predict the next visited URL, effectively framing the task as next-item prediction with URL identifiers instead of item ids. We use a hidden size of 192, 6 layers, and 4 attention heads.

The second variant (\textit{Next product}) is trained on product data to predict the next product-related interaction. Each event in this case consists of multiple categorical fields: event\_type, category, and price. The model predicts each of these fields using separate output heads. We use a hidden size of 128, 4 layers, and 4 attention heads.

\subsubsection{LM-based embeddings}

We also employ a pre-trained language model as a feature extractor \cite{llm4es}. All user interactions are converted into plain text, with one event per line. Apart from the event type and timestamp -- which we record for every event -- we include SKU, category, price, and quantized item name for "add to cart", "remove from cart", and "product buy" events; encoded URL for "page visit" events; and quantized query for "search query" events. Since the context size of the model is limited, we consider only the most recent 64 interactions. We then embed these textualized event sequences using SmolLM2-135M\footnote{\url{https://huggingface.co/HuggingFaceTB/SmolLM2-135M}}~\cite{allal2025smollm2}. Hereafter, we refer to these embeddings as "\textit{SmolLM2 all interactions}".

Additionally, we apply a similar preprocessing strategy exclusively to "search query" events. Given the greater homogeneity of the data, we utilize a tabular format and truncate each user’s history to the most recent 90 events. These embeddings are hereafter referred to as "\textit{SmolLM2 search queries}".

\subsubsection{Handcrafted features}

We build several features from logs of interactions with products and page visits. Firstly, we examine interaction patterns across multiple time windows, including the most recent week, the past month, and the complete user history. This analysis involves calculating the absolute number of interactions within each period, the relative change in interaction counts between consecutive periods such as week-over-week differences, and the average time interval between successive interactions. To build more informative features, we compute aggregated statistics capturing the variety of products and pages each user engaged with, the price distribution of interacted products, and the abandonment rate of cart items. 

\subsection{Combining All Together}

It is well known that ensemble methods generally outperform single models~\cite{caruana2004ensemble,dietterich2000ensemble}. Therefore, our final solution combines all of the previously described techniques. First, to keep the size of the final representation manageable, we apply dimensionality reduction using PCA to some of the individual embeddings. Specifically, we retain 64 components for \textit{SmolLM search queries}, 128 for \textit{SmolLM all interactions}, and 64 for all \textit{GRU-AE} variants except the \textit{day\_event\_type} model.
Next, since the individual embeddings come from different sources and have varying scales and distributions, we normalize each of them separately to improve the performance of downstream MLP models. We evaluated multiple scaling strategies and found that normalizing each user vector to unit length was most effective for model-generated representations.
For handcrafted features, we apply a quantile transformer\footnote{\url{https://scikit-learn.org/stable/modules/generated/sklearn.preprocessing.QuantileTransformer.html}} to map them to a uniform distribution. Finally, we concatenate all embeddings into 1066-dimensional vectors and handle missing values using mean imputation, which we found to outperform zero imputation.

%% file: content/tables/leaderboard.tex
\begin{table*}[htbp]
  \caption{\textbf{Leaderboard scores with top solutions and results of our single models.}}
  \label{tab:leaderboard}
  \centering
  \setlength{\tabcolsep}{4pt}
  
  \resizebox{0.82\textwidth}{!}{%
      \begin{tabular}{p{0.9cm} l *{8}{c}}
        \toprule
        & \textbf{Approach} &
          \makecell{\textbf{Churn}\\\textbf{Pred.}} &
          \makecell{\textbf{Category}\\\textbf{Prop.}} &
          \makecell{\textbf{Product}\\\textbf{Prop.}} &
          \textbf{Hidden 1} & \textbf{Hidden 2} & \textbf{Hidden 3} &
          \textbf{Sum} &
          \makecell{\textbf{Borda}\\\textbf{Count}} \\
        \midrule
    
        \multirow{5}{*}{\rotatebox[origin=c]{90}{\makecell{\textit{Top}\\\textit{Solutions}}}}
          & rec2                 & 0.7375 & 0.8179 & 0.8224 & 0.7717 & 0.8293 & 0.8161 & 4.7949 & 662 \\
        & ai\_lab\_recsys       & 0.7376 & 0.8180 & 0.8130 & 0.7649 & 0.8052 & 0.8117 & 4.7504 & 653 \\
        & SenseLab              & 0.7374 & 0.8180 & 0.8081 & 0.7637 & 0.8081 & 0.8116 & 4.7469 & 647 \\
        & ambitious             & 0.7389 & 0.8151 & 0.8041 & 0.7699 & 0.7948 & 0.8147 & 4.7375 & 641 \\
        & embednbreakfast       & 0.7366 & 0.8162 & 0.8134 & 0.7575 & 0.8039 & 0.8067 & 4.7343 & 638 \\
        \midrule
    
        \multirow{6}{*}{\rotatebox[origin=c]{90}{\makecell{\textit{Our Single}\\\textit{Models}}}}
          & GRU-AE week\_all         & 0.7288 & 0.8023 & 0.8007 & 0.7490 & 0.7476 & 0.8020 & 4.6304 & -- \\
        & GRU-AE day\_event\_type   & 0.7251 & 0.7828 & 0.7734 & 0.7376 & 0.7908 & 0.7888 & 4.5985 & -- \\
        & Next URL                  & 0.6869 & 0.7699 & 0.7613 & 0.6981 & 0.7468 & 0.7649 & 4.4279 & -- \\
        & Next product              & 0.6886 & 0.7508 & 0.7481 & 0.6555 & 0.7216 & 0.7427 & 4.3073 & -- \\
        & iALS page visits          & 0.6638 & 0.7334 & 0.7437 & 0.6705 & 0.7395 & 0.7429 & 4.2938 & -- \\
        & iALS categories           & 0.6807 & 0.7265 & 0.7309 & 0.6155 & 0.7174 & 0.7152 & 4.1862 & -- \\
        \bottomrule
      \end{tabular}%
   }
\end{table*}

%% file: content/4_results.tex
\input{content/tables/validation}

Table~\ref{tab:validation} presents the results of our local validation. Models based on the GRU autoencoder consistently outperform other approaches. The best option is \textit{GRU-AE week\_all}. Interestingly, a much simpler version that reconstructs only the day index and event\_type (\textit{GRU-AE day\_event\_type}) also shows competitive performance. Among the alternative models, the LM-based embeddings trained on all interactions (\textit{SmolLM2 all interactions}) achieve the second-best performance overall. Combining all embeddings into an ensemble significantly improves performance over any single approach, confirming the complementary nature of the learned representations.

Table~\ref{tab:leaderboard} shows the final leaderboard scores for the top 5 teams and results for several single models. The second row corresponds to our final ensemble. While our best GRU autoencoder model (\textit{GRU-AE week\_all}) is close to the top-performing solutions on most tasks, its performance on the hidden2 task lag noticeably behind. In contrast, the simpler \textit{GRU-AE day\_event\_type} model achieves excellent performance specifically on this task. In general, we observed that the hidden2 task was quite unstable and particularly sensitive to small variations in user embeddings, often resulting in large fluctuations in evaluation score.

%% file: content/tables/validation.tex
\begin{table}[htbp]
  \caption{Validation metrics on three open tasks and total sum of scores. Scores in bold denote the best single model for each task, while underlined scores denote the second best.}
  \label{tab:validation}
  \centering
  \setlength{\tabcolsep}{4pt}
  \resizebox{0.8\columnwidth}{!}{%
  
  \begin{tabular}{lcccc}
    \toprule
    \textbf{Approach} &
    \makecell{\textbf{Churn}\\\textbf{Pred.}} &
    \makecell{\textbf{Category}\\\textbf{Prop.}} &
    \makecell{\textbf{Product}\\\textbf{Prop.}} &
    \textbf{Sum} \\
    \midrule
    GRU\textendash AE week\_all           & \underline{0.8085} & \textbf{0.8007} & \textbf{0.7861} & 2.3953 \\
    GRU\textendash AE day\_event\_type    & \textbf{0.8132} & 0.7783 & \underline{0.7718} & 2.3633 \\
    GRU\textendash AE all                 & 0.8002 & \underline{0.7856} & 0.7638 & 2.3497 \\
    GRU\textendash AE sku\_text           & 0.6690 & 0.7565 & 0.7385 & 2.1640 \\
    \midrule
    SmolLM2 all interactions              & 0.7945 & 0.7608 & 0.7424 & 2.2977 \\
    Next product                          & 0.7428 & 0.7804 & 0.7409 & 2.2641 \\
    Next URL                              & 0.7332 & 0.7829 & 0.7478 & 2.2639 \\
    Handcrafted features                  & 0.7990 & 0.7428 & 0.6965 & 2.2384 \\
    iALS page visits                      & 0.7241 & 0.7468 & 0.7242 & 2.1951 \\
    iALS categories                       & 0.6594 & 0.7701 & 0.7482 & 2.1777 \\
    SmolLM2 search queries                & 0.6208 & 0.7207 & 0.7098 & 2.0513 \\
    LightFM                               & 0.6232 & 0.7135 &       0.6848 & 2.0216    \\
    \midrule
    Ensemble                              & 0.8212 & 0.8133 &       0.7912 & 2.4257 \\
    \bottomrule
  \end{tabular}
  }
\end{table}

%% file: content/5_conclusion.tex
In this work, we presented a sequence autoencoder approach for building Universal Behavioral Profiles from raw user interaction logs. By training the model to reconstruct the full sequence of user events, we encouraged the learned embeddings to capture essential behavioral patterns in a task-agnostic way. We also explored several complementary embedding strategies and demonstrated that combining them in an ensemble improves generalization across diverse downstream tasks. Our final solution achieved second place in the RecSys Challenge 2025, highlighting the effectiveness of sequence reconstruction as a foundation for universal user modeling.